\documentclass[twocolumn,pra,showpacs,preprintnumbers]{revtex4-1}%

\usepackage{booktabs}
\usepackage{amsmath, amssymb, amsfonts}
\usepackage{bm}
\usepackage{dsfont}
\usepackage{soul}
\usepackage{graphicx}
\usepackage{epsfig}
\usepackage{subfigure}
\usepackage{multirow}
\usepackage{makecell}
\usepackage[english]{babel}
\usepackage{mathrsfs}
\usepackage{lipsum}
\usepackage{appendix}
\usepackage{courier}
\usepackage[colorlinks,linkcolor=blue,citecolor=blue]{hyperref}%
\providecommand{\U}[1]{\protect \rule{.1in}{.1in}}
\begin{document}

\title{Measurement-induced generation of Schr\"{o}dinger cat states in cavity QED}
\author{Tong Wang$^{1}$}
\author{Peng-Fei Wei$^{1}$}
\author{Hai-Jun Xing$^{1}$}
\author{Zhihai Wang$^{1}$}
\email{wangzh761@nenu.edu.cn}
\affiliation{1. Center for Quantum Sciences and School of Physics, Northeast Normal University, Changchun 130024, China}
\begin{abstract}

Schr\"{o}dinger cat states, representing coherent superpositions of macroscopically distinguishable states, are indispensable nonclassical resources for continuous-variable quantum information processing. Existing generation protocols typically rely on strong nonlinear interactions, complicated control techniques, or engineered dissipation, posing challenges for experimental implementation. Here, we propose a simple measurement-based protocol for generating Schr\"{o}dinger cat states in a cavity-QED system by combining coherent driving, dispersive atom--cavity interactions, and atomic postselection. The atom--cavity interaction establishes coherent correlations between the atomic and photonic degrees of freedom, while the subsequent atomic postselection projects the cavity field onto a non-Gaussian superposition state with pronounced Wigner negativity. Numerical simulations based on the Lindblad master equation show that the generated Schr\"{o}dinger cat states remain robust against moderate cavity dissipation. Our results demonstrate that conditional atomic measurements provide an effective and experimentally accessible approach for preparing nonclassical cavity states without relying on strong optical nonlinearities or engineered dissipation.
\end{abstract}

\maketitle

\section{Introduction}

Schr\"{o}dinger cat states, namely coherent superpositions of macroscopically distinguishable classical states, are among the most representative nonclassical resources in continuous-variable quantum information processing \cite{PhysRevA456570,PhysRevLett5713,ng3tp1n1,He2023,Zheng2026PRL}. Owing to their wide applications in quantum computation \cite{PhysRevA65042305,PhysRevA68042319,Ofek2016,Lund2008,PhysRevA592631,PhysRevLett111120501,Mirrahimi2014,PhysRevA101063802,
PhysRevA105013504,science1243289,Hu2019,Chamberland2022}, quantum communication \cite{RevModPhys77513,Gilchrist2004,PhysRevA64052308,Deleglise2008Nature,Neergaard2013},
 and quantum metrology \cite{PhysRevA66023819,PhysRevLett107083601,
 scienceadf7553,PhysRevA106022619,Tatsuta2019,Huang2015,xing2026}, the generation of Schr\"{o}dinger cat states has attracted extensive theoretical and experimental interest. A variety of physical platforms, including cavity quantum electrodynamics (QED), trapped ions, superconducting circuits, and optical systems, have been explored for their realization \cite{scienceadf7553}. Existing approaches mainly rely on strong Kerr nonlinearities \cite{AMiranowicz1990,He2023,PhysRevA101063802,PhysRevA105013504}, photon subtraction \cite{PhysRevA553184,PhysRevA103013710,endo2025highratephotonsubtractionsqueezed,Ourjoumtsev2007,Neergaard2006,Takahashi2008},
or reservoir engineering \cite{PhysRevA106023714,PhysRevA104013715,Mirrahimi2014,Leghtas2015Science}. Although remarkable progress has been achieved, these methods generally require strong nonlinear interactions, complicated control techniques, or carefully engineered dissipation, motivating the search for experimentally accessible alternatives for generating nonclassical continuous-variable states \cite{Deleglise2008Nature}.

Recently, measurement-induced quantum state engineering has emerged as a powerful route toward the preparation of nonclassical states
\cite{deGraaf2025npjQI,PhysRevA73041801,
Sun2014,Deng2026Optica,PhysRevA61032302}. Since the dynamics of most continuous-variable systems are governed by Gaussian Hamiltonians, Gaussian unitary evolution alone cannot generate non-Gaussian states from Gaussian inputs \cite{PhysRevLett89137903,RevModPhys84621}. Consequently, conditional measurements have become an indispensable tool for generating non-Gaussianity, enabling the preparation of Schr\"{o}dinger cat states and other nonclassical continuous-variable states \cite{Sun2014,Yan2026PRA}. These developments naturally motivate the exploration of simple and experimentally feasible measurement-based protocols for Schr\"{o}dinger cat state generation.

In this work, we propose a conditional state-preparation protocol for generating Schr\"{o}dinger cat states in a cavity-QED system. The protocol combines coherent driving of a cavity mode with dispersive interactions between the cavity field and a sequence of flying two-level atoms \cite{Brune1992PRA,RevModPhys73565,Guerlin2007,PhysRevA100053825}. During each interaction cycle, the cavity field becomes entangled with the atomic internal state, and a subsequent postselection on the atom projects the cavity into a non-Gaussian superposition state \cite{PhysRevA73041801}. The underlying physical mechanism can be understood from the perspective of quantum erasure \cite{PhysRevA252208,Scully1991,Liu2017SciAdv}. Specifically, the dispersive interaction first encodes the which-way information \cite{Scully1991} associated with different coherent-state branches into the atomic state, whereas the subsequent postselection erases this information and restores coherent interference between the branches, thereby generating the Schr\"{o}dinger cat state. The protocol can also be naturally generalized to multi-component Schr\"{o}dinger cat states through successive interaction cycles.

We first present analytical results for the ideal closed-system case and then investigate the influence of cavity dissipation by solving the Lindblad master equation. Numerical simulations show that the generated Schr\"{o}dinger cat states remain robust against moderate photon loss and retain pronounced Wigner negativity, demonstrating their nonclassical character. Our work provides a simple and experimentally accessible approach for measurement-induced continuous-variable quantum state engineering in cavity-QED systems.
\section{MODEL AND HAMILTONIAN}
\begin{figure}[t]
    \centering
    \includegraphics[width=0.9\linewidth]{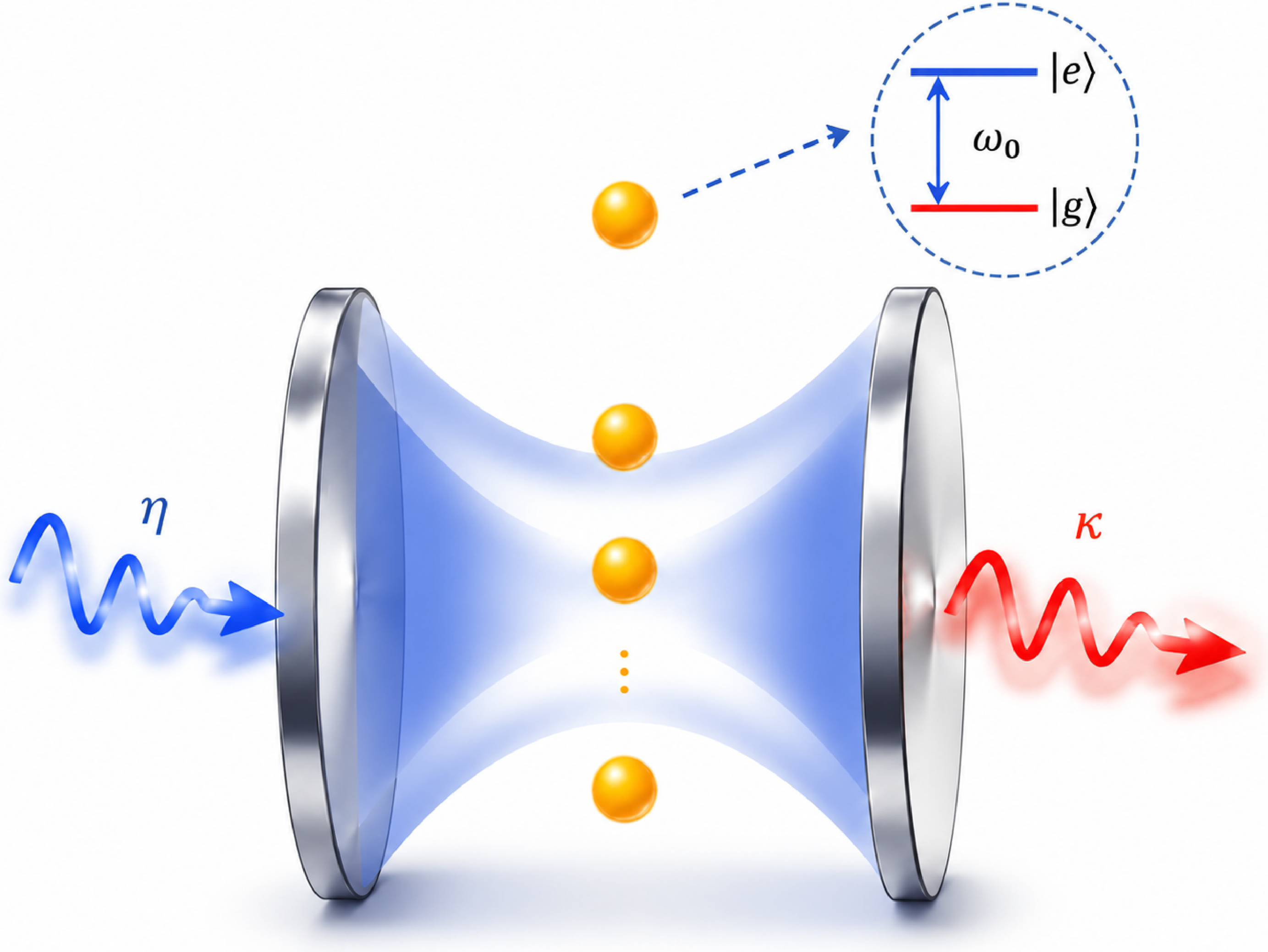}
    \caption{A schematic diagram of the model used for generating the Schr\"{o}dinger cat state is shown in the figure. A series of two-level atoms which are prepared in the same initial state pass through a single-mode cavity one by one.}
    \label{fig:placeholder}
\end{figure}

The cavity-QED setup considered here is illustrated schematically in Fig.~\ref{fig:placeholder}. A single-mode cavity with resonance frequency $\omega_a$ is coherently driven by a classical field of amplitude $\eta$ and frequency $\omega_c$. Under resonant driving ($\omega_c=\omega_a$), the driving Hamiltonian in the rotating frame is $H=\eta(a^\dagger+a)$, where $a$ is the annihilation operator of the cavity mode. A sequence of two-level atoms then traverses the cavity and interacts dispersively with the intracavity field. In the large atom--cavity detuning regime, the interaction is described by $H_{\rm int}=\chi a^\dagger a\sigma_z$, where $\chi$ denotes the effective dispersive coupling strength \cite{PhysRevA69062320,Brune1992PRA}. Such dispersive atom--cavity interactions have been extensively employed in cavity-QED experiments for quantum state manipulation and nondestructive quantum measurements \cite{Krastanov2015PRA,RevModPhys73565,Guerlin2007}.

In this paper, we propose a periodic protocol for generating bosonic Schr\"{o}dinger cat states with a cycle period $T=\tau_1+\tau_2$. During the interval $t\in(nT,nT+\tau_1)$, where $n$ is an integer, the atom remains outside the cavity and the cavity mode is driven resonantly by the classical field. During the subsequent interval $t\in(nT+\tau_1,(n+1)T)$, the driving field is switched off, and the atom enters the cavity and interacts dispersively with the cavity mode. Repeating this sequence yields the time-dependent Hamiltonian
\begin{equation}
H=\left\{
\begin{aligned}
&\eta(a^{\dagger}+a),\quad nT< t< nT+\tau_1,\\
&\chi a^{\dagger}a\sigma_{z},\quad\quad nT+\tau_1\leq t< (n+1)T.
\end{aligned}
\right.
\label{eqH}
\end{equation}

To account for photon loss, we describe the dynamics of the system by the Lindblad master equation \cite{Lindblad1976,Carmichael1993}

\begin{equation}
\frac{d\rho}{dt}=-i[H,\rho]+\kappa\left(2a\rho a^\dagger-a^\dagger a\rho-\rho a^\dagger a\right),
\label{master}
\end{equation}
where $\kappa$ denotes the cavity decay rate.

We now show how postselection of the outgoing atom enables the conditional generation of Schr\"{o}dinger cat states in the cavity. We first present an analytical treatment in the ideal lossless limit and then examine the influence of cavity dissipation on the state-preparation protocol.

\section{ANALYTICAL RESULTS for THE CLOSED SYSTEM}
We first consider the ideal case in which cavity dissipation is neglected. In this limit, the atom--cavity system evolves unitarily. Since the cavity dynamics consists solely of coherent driving and dispersive interactions, the evolution remains Gaussian and therefore cannot generate Schr\"{o}dinger cat states from an initial Gaussian state \cite{PhysRevLett89137903,RevModPhys84621}. To overcome this limitation, we perform projective postselection on each atom after it exits the cavity, thereby inducing the required non-Gaussian operation.

Initially, the cavity is prepared in the vacuum state, $|\psi_c(0)\rangle=|0\rangle$. During the first stage, $0<t<\tau_1$, the resonant driving field displaces the cavity field into the coherent state

\begin{equation}
|\psi_c(\tau_1)\rangle
=e^{-i\eta(a^\dagger+a)\tau_1}|0\rangle
=|-i\eta\tau_1\rangle.
\end{equation}

Next, the external driving is switched off and a two-level atom, initially prepared in the coherent superposition state
$|\psi_a\rangle=\cos\theta_0|e\rangle+\sin\theta_0|g\rangle$,
is injected into the cavity. At the end of the interaction, i.e., at $t=T$, the atom--cavity system evolves into
\begin{equation}
\begin{aligned}
|\phi_{ac}(T)\rangle=&e^{-i\chi a^\dagger a\sigma_z\tau_2}|\psi_a;\psi_c(\tau_1)\rangle\\
=&\cos\theta_0|e;-i\eta\tau_1
e^{-i\chi\tau_2}\rangle+\sin\theta_0|g;-i\eta\tau_1
e^{i\chi\tau_2}\rangle.
  \label{equaandc}
\end{aligned}
\end{equation}

Eq.~(\ref{equaandc}) shows that the dispersive atom--cavity interaction splits the initial coherent state into two coherent-state branches with opposite phase rotations, each correlated with a different atomic state. As a result, the atom becomes entangled with the cavity field. A projective measurement in the $\{|e\rangle,|g\rangle\}$ basis preserves the which-way information encoded in the atomic state and therefore projects the cavity field onto only one of the two coherent-state branches. Such a measurement alone cannot generate a Schr\"{o}dinger cat state.

To erase the which-way information and recover the quantum interference between the two branches, we instead perform postselection on the atom \cite{Scully1991}. Without loss of generality, the post-selected state is chosen as $|\psi_f\rangle=\cos\theta_f|e\rangle+\sin\theta_f|g\rangle$, and, for simplicity, we set $\theta_f=\theta_0=\theta$. Conditioned on a successful postselection, the cavity state becomes
\begin{equation}
\begin{aligned}
|\psi_{c,1}\rangle
&=\frac{1}{\sqrt{P_1}}\langle\psi_f|\phi_{ac}(T)\rangle\\
&=\frac{1}{\sqrt{P_1}}\Big(
\cos^2\theta|-i\eta\tau_1 e^{-i\chi\tau_2}\rangle
+\sin^2\theta|-i\eta\tau_1 e^{i\chi\tau_2}\rangle
\Big),
\label{liangjiao}
\end{aligned}
\end{equation}
which represents a Schr\"{o}dinger cat state, namely, a coherent superposition of two phase-separated coherent states. Here, $P_1$ is the normalization constant, which also gives the success probability of the postselection.
\begin{figure*}[!t]
    \centering

    \includegraphics[trim=4.2cm 3.2cm 1cm 3cm, clip, width=1.13\textwidth]{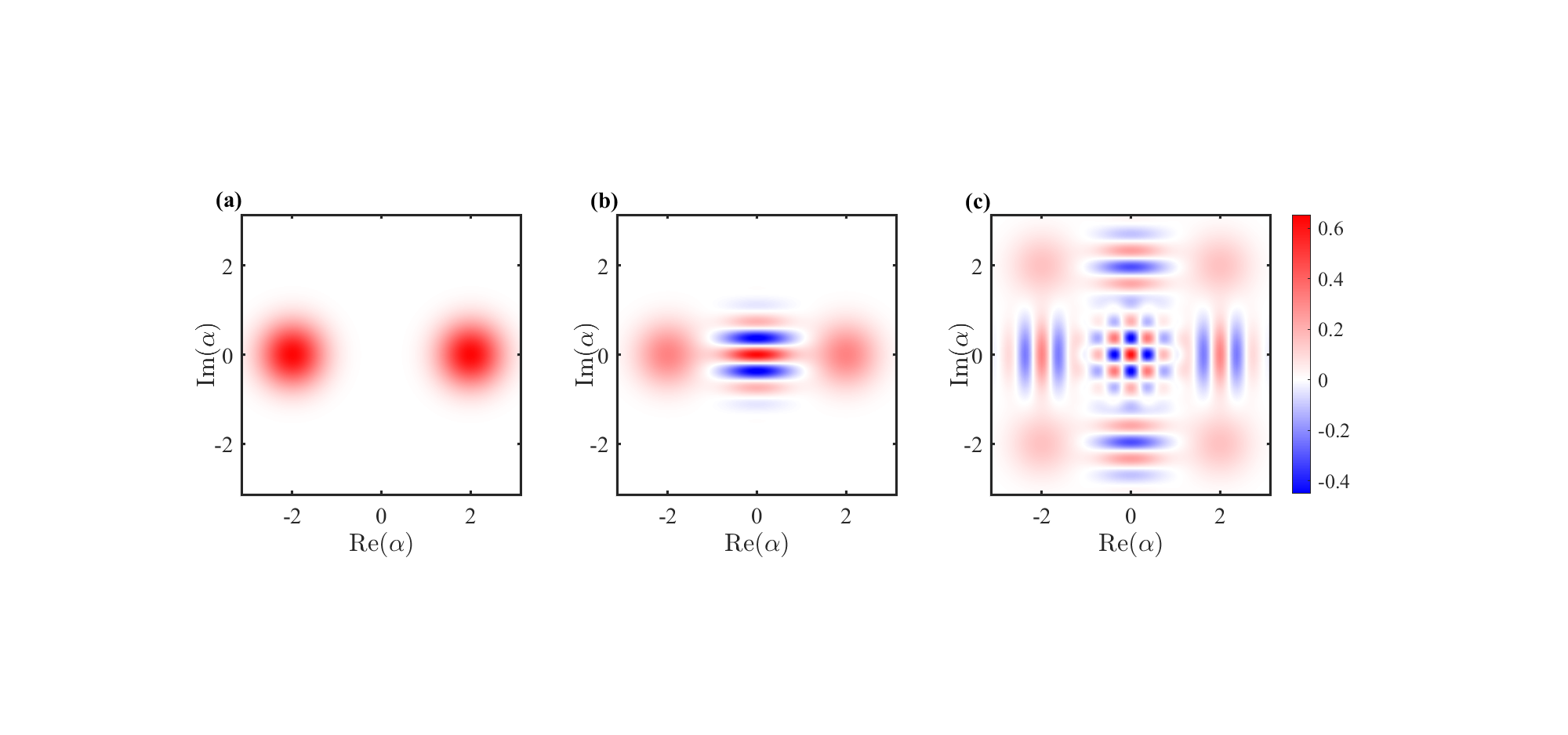}
    \caption{Wigner functions of Schr\"{o}dinger cat states without dissipation. (a) The two-legged Schr\"{o}dinger cat state. (b) The four-legged Schr\"{o}dinger cat state. The drive area and interaction phase are set as $\eta \tau_1 = 2$ and $\chi \tau_2 = \pi/2$, respectively, with the remaining parameters chosen as $\theta_0 = \theta_f = \pi/4$.}
    \label{fig2}
\end{figure*}
To characterize the nonclassical properties of the generated state, we calculate its Wigner function in phase space, defined as
\begin{equation}
W(\alpha)=\frac{2}{\pi}\mathrm{Tr}\left[\hat{\rho}_c\hat{D}(\alpha)\hat{\Pi}\hat{D}^{\dagger}(\alpha)\right],
\end{equation}
where $\hat{\rho}_c$ is the density matrix of the cavity field, $\hat{D}(\alpha)$ is the displacement operator, and $ \hat{\Pi}=(-1)^{\hat{a}^{\dagger}\hat{a}}$ is the photon-number parity operator \cite{PhysRevA15449,Barnett1997}.

With the projective measurement in the atomic energy basis,
$E_1=|e\rangle\langle e|$ and $E_2=|g\rangle\langle g|$,
the atomic which-way information is preserved and the cavity field collapses into one of the two coherent-state branches.
The corresponding Wigner function is shown in Fig.~\ref{fig2}(a).
Two Gaussian wave packets are clearly observed, centered at
$-i\eta\tau_1e^{-i\chi\tau_2}$ and
$-i\eta\tau_1e^{i\chi\tau_2}$, respectively.
However, no interference fringes appear between the two components, and the Wigner function remains nonnegative, indicating the absence of quantum coherence between the two coherent-state branches.

In contrast, when the atom is postselected onto the superposition state
$|\psi_f\rangle$, the corresponding Wigner function is shown in Fig.~\ref{fig2}(b).
Pronounced interference fringes emerge in the phase-space region between the two coherent-state components, accompanied by negative values of the Wigner function.
These interference fringes, rather than the mere coexistence of two coherent-state components, provide the characteristic signature of a Schr\"{o}dinger cat state.
Physically, the post-selection projects the atom onto a superposition basis and erases the which-way information stored in the atomic degree of freedom, thereby restoring the coherent interference between the two cavity-field branches.
The resulting Wigner negativity confirms the nonclassical nature of the generated cavity state.

The proposed protocol can be naturally extended to generate multi-component Schr\"{o}dinger cat states by repeating the driving--interaction--postselection cycle. As the simplest example, after a second cycle consisting of resonant driving during $(T,T+\tau_1)$, dispersive atom--cavity interaction during $(T+\tau_1,T+\tau_2)$, and the subsequent atomic postselection, the cavity state becomes
\begin{equation}
\begin{aligned}
|\psi_{c,2}\rangle
=\frac{1}{\sqrt{P_2}}\Big[
&\cos^4\theta e^{i\varphi}|-i\eta\tau_1 e^{-i\chi\tau_2}-i\eta\tau_1e^{-i\chi\tau_2}\rangle \\
&+\frac{1}{4}\sin^4(2\theta)e^{i\varphi}|-i\eta\tau_1 e^{-i\chi\tau_2}-i\eta\tau_1 e^{i\chi\tau_2}\rangle \\
&+\frac{1}{4}\sin^4(2\theta)e^{-i\varphi}|-i\eta\tau_1 e^{i\chi\tau_2}-i\eta\tau_1 e^{-i\chi\tau_2}\rangle \\
&+\sin^4\theta e^{-i\varphi}|-i\eta\tau_1 e^{i\chi\tau_2}-i\eta\tau_1e^{i\chi\tau_2}\rangle \Big],
\end{aligned}
\end{equation}
where $\varphi = \eta^2{\tau_1}^2\sin(\chi\tau_2)$.
Figure~\ref{fig2}(c) shows the corresponding Wigner function. Compared with the two-component cat state in Fig.~\ref{fig2}(b), the cavity field now consists of four coherent-state components distributed in phase space. More significantly, the interference pattern evolves from one-dimensional fringes into a two-dimensional interference network, reflecting the coherent interference among all four components. The checkerboard-like structure with alternating positive and negative values in the central region provides clear evidence of the enhanced quantum coherence associated with the four-component Schr\"{o}dinger cat state \cite{Zurek2001}.

More interestingly, the above protocol admits a recursive formulation, which immediately demonstrates its scalability for generating multi-component Schr\"{o}dinger cat states. After $N$ interaction cycles, the cavity field evolves into a Schr\"{o}dinger cat state consisting of $2^N$ coherent-state components. To show this explicitly, we assume that after the $n$th cycle the cavity state can be written as
\begin{equation}
|\psi_{c,n}\rangle=\sum_{i=1}^{2^n}C_i|\beta_i\rangle,
\label{SN}
\end{equation}
where $|\beta_i\rangle$ denotes a coherent-state component with complex amplitude $\beta_i$. After one additional driving--interaction--postselection cycle, the cavity state is updated according to the recursive relation
\begin{equation}
\begin{aligned}
|\psi_{c,n+1}\rangle
=&
\sum_{i=1}^{2^n}
C_i\xi_i
\cos^2\theta
|(\beta_i-i\eta\tau_1)e^{-i\chi\tau_2}\rangle
\\
&+
\sum_{i=1}^{2^n}
C_i\xi_i
\sin^2\theta
|(\beta_i-i\eta\tau_1)e^{i\chi\tau_2}\rangle .
\end{aligned}
\label{SNp1}
\end{equation}
where $\xi_i=\exp[-i\eta\tau_1{\rm Re}(\beta_i)]$ is the phase accumulated during the driving stage. Equations~(\ref{SN}) and (\ref{SNp1}) show that each coherent-state component generated in the $n$th cycle is deterministically split into two new components in the subsequent cycle. Consequently, the number of coherent-state components doubles after every cycle, leading to the systematic generation of $2^N$-component Schr\"{o}dinger cat states.

\begin{figure}[htbp]
    \centering
    \hspace*{-0.55cm}
   \includegraphics[trim=1cm 0.5cm 0cm 0.25cm, clip,width=1.28\linewidth]{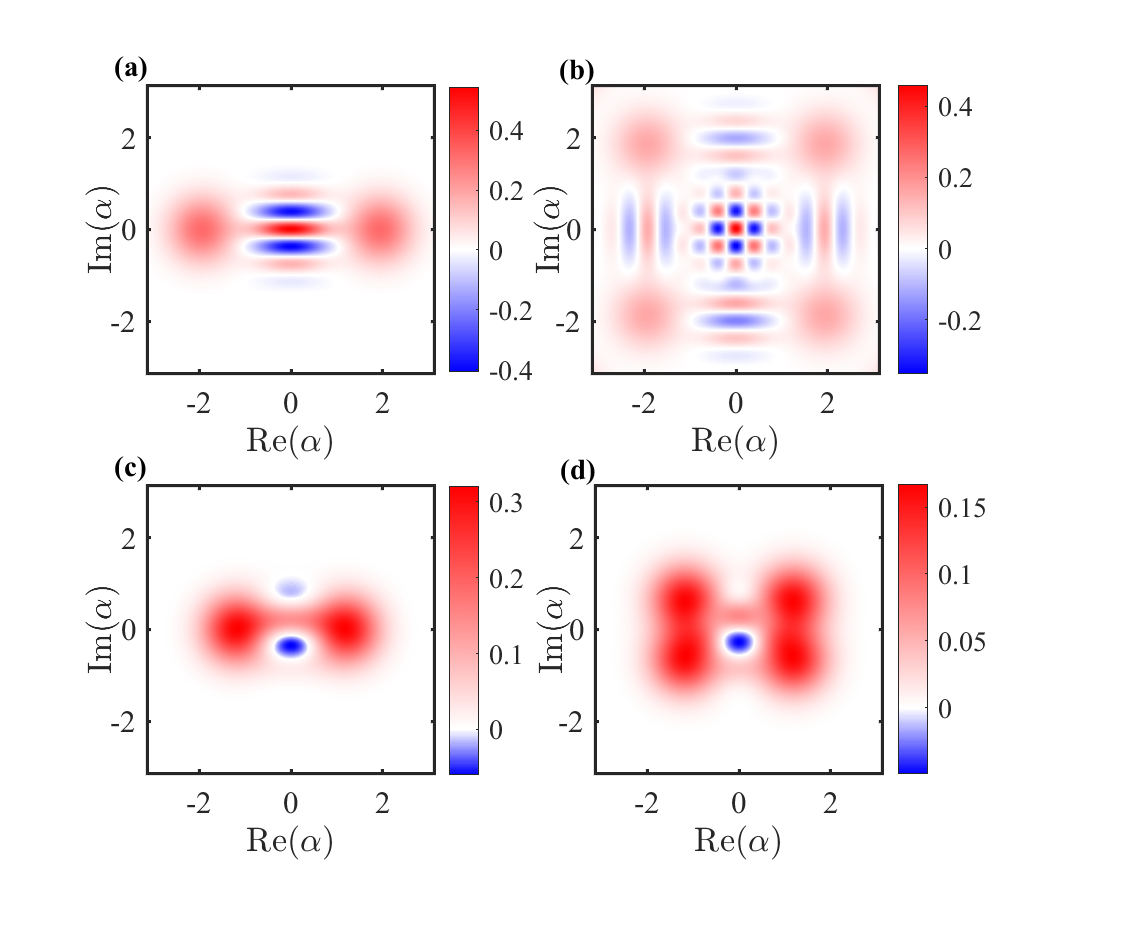}
    \caption{Influence of dissipation on the Wigner functions of Schr\"{o}dinger cat states. Two-legged and four-legged Schr\"{o}dinger cat states are shown under (a, b) weak dissipation ($\kappa = 0.01\eta$) and (c, d) stronger dissipation ($\kappa = 0.075\eta$). The drive area and interaction phase are set as $\eta \tau_1 = 2$ and $\chi \tau_2 = \pi/2$, respectively, with the remaining parameters chosen as $\theta_0 = \theta_f = \pi/4$.}
    \label{fig3}
\end{figure}

\section{Numerical results for the dissipative cavity}
In the presence of cavity dissipation, the recursive analytical solution derived for the closed system is no longer valid. We therefore solve the Lindblad master equation numerically to investigate the robustness of the protocol,  to evaluate the robustness of the state-preparation protocol against photon loss.

The cavity is initially prepared in the vacuum state. After one combination of driving and interaction, the density matrix of the atom--cavity system, denoted by $\rho_{ac}$, is obtained by numerically solving Eq.~(\ref{master}). Projecting the atomic state onto the post-selected state yields the reduced density matrix of the cavity, $\rho_{c}={\rm Tr}_a(\mathcal{P}_f\rho_{ac}\mathcal{P}_f)$, where $\mathcal{P}_f=|\psi_f\rangle\langle\psi_f|\otimes I_c$, with $I_c$ the identity operator acting on the cavity Hilbert space. The resulting cavity state corresponds to the two-component Schr\"{o}dinger cat state. Repeating one additional driving--interaction--postselection cycle generates the four-component cat state in the dissipative regime. In the limit of vanishing cavity decay ($\kappa=0$), the numerical results recover the analytical solutions obtained in the previous section. Figure~\ref{fig3} presents the corresponding Wigner functions, illustrating how cavity dissipation influences the generated Schr\"{o}dinger cat states.

We first consider the weak-dissipation regime with $\kappa=0.01\eta$. The corresponding Wigner functions of the two- and four-component Schr\"{o}dinger cat states are shown in Figs.~\ref{fig3}(a) and (b), respectively. Compared with the ideal results in Figs.~\ref{fig2}(b) and (c), the overall phase-space structures are well preserved. The coherent-state components remain clearly distinguishable, while the interference fringes between them are only slightly reduced. In particular, the negative regions of the Wigner function, which are the hallmark of quantum coherence and nonclassicality, remain clearly visible. This demonstrates that the quantum coherence generated by the postselection survives under weak cavity dissipation.

As the dissipation increases to $\kappa=0.075\eta$, the Wigner functions shown in Figs.~\ref{fig3}(c) and (d) reveal that the generated states still retain nonclassical features, as evidenced by the remaining negative regions of the Wigner function. However, the quantum coherence is substantially degraded compared with the ideal case \cite{Deleglise2008Nature}.

For the two-component Schr\"{o}dinger cat state [Fig.~\ref{fig3}(c)], the coherent-state components move closer to the origin owing to photon loss, while the interference pattern becomes markedly asymmetric with respect to the ${\rm Im}(\alpha)$ axis. In particular, the negative part of the Wigner function is more pronounced in the region ${\rm Im}(\alpha)<0$ than in ${\rm Im}(\alpha)>0$, indicating that cavity dissipation suppresses the quantum coherence in a direction-dependent manner.

A similar but more pronounced behavior is observed for the four-component Schr\"{o}dinger cat state [Fig.~\ref{fig3}(d)]. The coherent-state components located in the upper and lower half planes become strongly compressed and nearly overlap, accompanied by a substantial suppression of the corresponding interference fringes. By contrast, the two components distributed along the ${\rm Re}(\alpha)$ axis remain well separated. Their interference survives predominantly in the region ${\rm Im}(\alpha)<0$, whereas it is almost completely suppressed in the opposite half plane. These results indicate that stronger cavity dissipation not only reduces the overall quantum coherence but also induces a pronounced asymmetry in the phase-space interference pattern. The observed asymmetry suggests that cavity dissipation affects different coherent-state branches unequally under the present driving protocol.

\section{Conclusion}
In conclusion, we have proposed a conditional protocol for generating Schr\"{o}dinger cat states in a cavity-QED system by combining coherent driving, dispersive atom--cavity interactions, and atomic postselection. Unlike conventional approaches relying on strong nonlinear interactions or engineered dissipation, the present scheme introduces the required non-Gaussianity through atomic postselection, providing a simple and experimentally accessible approach for preparing nonclassical cavity states.

The underlying physical mechanism can be understood from the perspective of quantum erasure. During the atom--cavity interaction, the which-way information associated with different coherent-state branches is encoded into the atomic internal state. The subsequent postselection erases this information and restores the coherent interference between different branches, thereby generating the Schr\"{o}dinger cat state. Furthermore, the protocol can be naturally generalized to multi-component Schr\"{o}dinger cat states through successive interaction cycles.

We have also investigated the influence of cavity dissipation by solving the Lindblad master equation. Numerical results show that the generated two- and four-component Schr\"{o}dinger cat states remain robust against moderate photon loss and preserve pronounced Wigner negativity, demonstrating the feasibility of the proposed protocol under realistic experimental conditions. We expect that the present work will provide a useful approach for preparing nonclassical continuous-variable states in cavity-QED systems and stimulate further studies on measurement-assisted quantum state engineering.

\section*{Acknowledgments}

This work is supported by the Natural Science Foundation of China (Grants No.~12375010), the Quantum Science and Technology-National Science and
Technology Major Project (No. 2023ZD0300700) and the Fundamental Research
 Funds for the Central Universities (Grant No.~2412023QD007).

\section*{DATA AVAILABILITY}

The data that support the findings of this article are not publicly available. The data are available from the authors upon reasonable request.

\end{document}